# Direct evidence and atomic-scale mechanisms of reduced dislocation mobility in ZnS under illumination

Mingqiang Li[a,#], Kun Luo[b,#], Xiumei Ma[c], Boran Kumral[d], Peng Gao[c], Tobin Filleter[d], Qi An[b*], Yu Zou[a*]

[a]Department of Materials Science and Engineering, University of Toronto, Toronto, ON M5S 3E4, Canada
[b]Department of Materials Science and Engineering, Iowa State University, Ames, IA 50011, USA
[c]Electron Microscopy Laboratory, School of Physics, Peking University, Beijing, 100871, China
[d]Department of Mechanical & Industrial Engineering, University of Toronto, Toronto, M5S 3G8, Canada

# These authors contribute equally to this work.
* Corresponding authors: mse.zou@utoronto.ca (Y.Z.); qan@iastate.edu (Q.A.)

## Abstract

Photo-plasticity in semiconductors, wherein their mechanical properties such as strength, hardness and ductility are influenced by light, has been reported for several decades. Although such phenomena have drawn significant attention to the manufacturability of deformable semiconductor devices, their underlying mechanisms are not well understood due to the lack of direct evidence. Here we provide experimental observation and atomic insights into the reduced mobility of dislocations in zinc sulfide (ZnS), as a model material, under light. Using photo-nanoindentation and transmission electron microscopy, we observe that dislocations glide shorter distances under light than in darkness, and there are no apparent deformation twins in both conditions. By atomic-scale simulations, we demonstrate that the decreased dislocation mobility is attributed to the increased Peierls stress and enhanced stress fields around dislocation cores due to photoexcitation. This study improves the understanding of photo-plastic effects in inorganic semiconductors, offering opportunities for modulating their mechanical properties using light.

Dislocations are linear crystal defects commonly found in bulk semiconductors and epitaxial semiconductor films, often arising from thermal strain or lattice mismatch [1-3]. These defects significantly influence their mechanical [4], electrical [5], optical [6], and thermal [7] properties. Extensive studies have investigated the manipulation of dislocations in semiconductors through external stimuli such as illumination, electric fields, and thermal annealing [8-11]. Previous findings suggest that light could affect the flow stress of many semiconductors, a phenomenon known as photo-plastic effect [8, 12-14]. For example, the strength of CdS is sensitive to visible light [15]: The compressive flow stress immediately increases under light, while the stress drops back once the light source is off. Similar effects of light on plastic deformation have been observed in II-VI (e.g., ZnS and CdTe) and III-V (e.g., GaAs and GaP) compounds and elemental semiconductors (e.g., Si and Ge) [8, 12, 16-20]. In ZnS, its Vickers microhardness is higher under UV illumination than that in darkness [21], and ~45 % plastic deformation of single-crystalline ZnS is achieved in darkness [22]. Such a photo-plastic effect is also sensitive to light intensity [23]: The hardness first increases and then becomes saturated in ZnS with increasing light intensity. Although previous reports suggest that the photo-plasticity phenomena are attributed to the changes of dislocation mobility under illumination and in darkness [13, 24], the evolution of dislocation under these conditions remains elusive.

The nanoscale size and buried feature of dislocations impede the identification and tracking of their evolutions, especially under illumination. One hypothesis suggests that illumination increases dislocation charge and introduces photoionized traps, thereby raising the Peierls barrier of dislocation glide [13, 25, 26]. Atomic simulations indicate that the reconstruction of dislocation cores induced by photo-induced charges may result in increasing the dislocation glide barriers [27, 28]. Other studies suggest that photo-induced carriers modify the energy landscape of the γ-surface, hardening the semiconductors [29, 30]. Moreover, under light, reductions in strain concentration at dislocation cores in semiconductors lead to a ductility-to-brittleness transition [31]. Although these simulations have provided valuable insights into understanding the photo-plastic effect, the direct experimental comparison of the dislocation evolution under light and darkness has not been well reported.

The nanoindentation method is suitable for studying the plastic deformation of semiconductors, because traditional mechanical testing, such as tensile and compression, is generally not suitable for brittle semiconductors [24, 32, 33]. Fig. 1a shows the schematics of nanoindentation on ZnS single crystals under light. Fig. 1b shows the representative nanoindentation load-depth curves of ZnS in darkness and under UV light, respectively, suggesting the light-induced hardening effect in ZnS. We provided the details of experiments and simulations in the supplementary materials. Fig. 1c shows that the average hardness of the ZnS samples increases from 2.49 GPa in darkness to 2.99 GPa under illumination, which is consistent with the hardening effect in literature reports [18, 21]. Multiple indentations were tested for various lighting conditions to ensure the repeatability of the measurements (Fig. S1 in the supplementary). The average modulus also increases from 87.56 GPa (in dark) to 97.26 GPa (under light), as shown in Fig. S2. No microcracks are observed in all indents as revealed by atomic force microscopy (AFM) (Fig. S3).

To further investigate the dependence of dislocation mobility on the light wavelength and intensity, we carried out photo-indentation measurements on ZnS under the light conditions with 281 nm, 365 nm, 800 nm, and 940 nm wavelengths in Fig. 1d. At each wavelength, we also studied the dependence of the photo-plastic effect on light intensity (Fig. S4), with light intensity measured using an optical power meter (Fig. S5). We calculate the change of hardness ($\Delta H$) under the maximum light intensity for each wavelength by the following equation:

$$\Delta H = \frac{Hardness_{Light} - Hardness_{dark}}{Hardness_{dark}} \times 100\% \qquad (1)$$

The maximum photo-plasticity effect occurs under 365 nm light since ZnS has the highest absorbance efficiency around 365 nm, which corresponds to its bandgap energy [34].

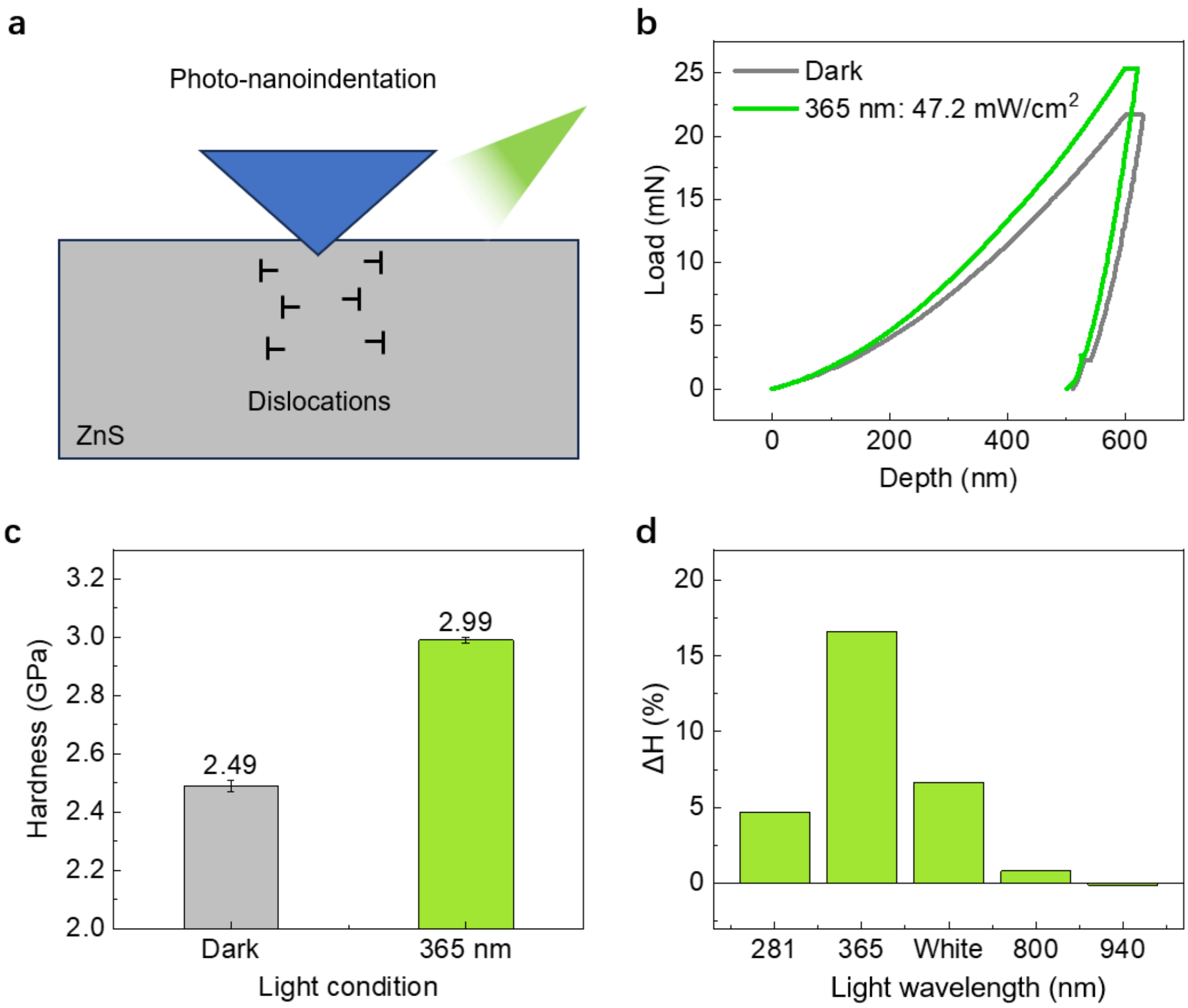

Fig. 1. Nanoindentation on ZnS in darkness and under light. (a) Schematics of the experimental setup; Dislocations are formed and propagated under the indents. (b) Typical load-depth curves in darkness (grey) and under illumination (green). The depth is set as 600 nm for both indents. (c) The hardness of ZnS in darkness and under illumination. (d) The change of hardness $\Delta H$ under each light wavelength.

The transmission electron microscopy (TEM) bright-field images show the distributions of dislocations underneath the indents in darkness and under light, respectively (Fig. 2 a-b). TEM samples from indents are prepared by FIB as shown in Supplementary Fig. S6. The enlarged TEM images of the regions ~5 μm from the surfaces are used to visualize dislocation distributions. At the bottom of the samples, the dislocation density for the light indent is lower than that for the dark indent (Fig. 2 a-b). To quantitatively compare the dislocation distributions, we counted the number of dislocations under the indents. Fig. 2c illustrates the number of dislocations as a function of the distance from the surface. The observation of decreased dislocation densities in both dark and light indents agrees with the previous characterization of dislocations near indents [35]. It shows that the dislocation density for the light indent is about 59 % less than that in the dark indent at the bottom of the samples, suggesting reduced dislocation mobility under illumination. In addition to the indented regions, TEM images and electron diffraction patterns of the undeformed surface do not show dislocations and splitting of diffraction spots (Fig. S7).

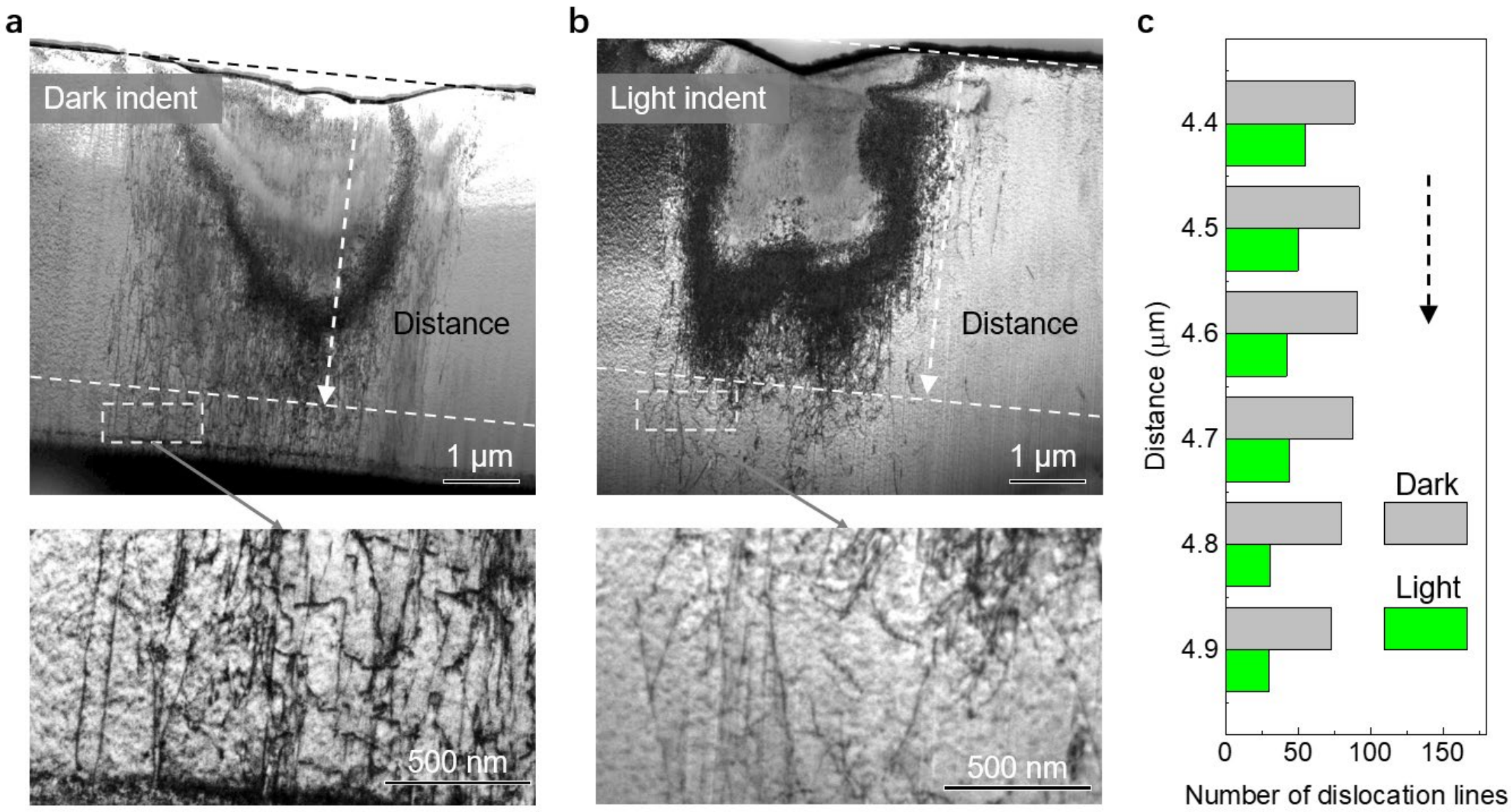

Fig. 2. Dislocation distributions underneath indents after nanoindentation in the darkness and under light. (a) An overview of dislocations underneath the dark indent. An enlarged view of the bottom area is provided to observe dislocation lines. (b) An overview of dislocations underneath the light indent. An enlarged view of the bottom area is provided to observe dislocation lines. (c) The number of dislocation lines as a function of distance away from the surface. The dislocation density in light indents is lower than those in dark indents.

Using the selected area electron diffractions (SAED), we identify the sample after the dark indentation exhibits an almost perfect single-crystal structure (Fig. 3 a-d), and no obvious twining structures are observed. Hence, dislocation motion is the main mechanism of plastic deformation in ZnS, as reported in the literature [24, 36]. The observed increase in hardness under illumination suggests a decrease in the dislocation mobility, due to the increase in the Peierls stress [37]. For the sample after the light indentation (Fig. 3e), we observe the splitting of diffraction spots as pointed out by the yellow arrows in Fig. 3 f and g, indicating the lattice rotation due to the high density of geometrically necessary dislocations (GNDs) near the indent, as schematically illustrated in Fig. 3e. Such lattice rotation due to GNDs is in good agreement with the Nix-Gao model [38-41] and a detailed explanation is provided in Fig. S8. The pronounced lattice rotation induced by high GNDs beneath the indents offers additional evidence that the dislocations propagate less extensively under illumination than those in darkness.

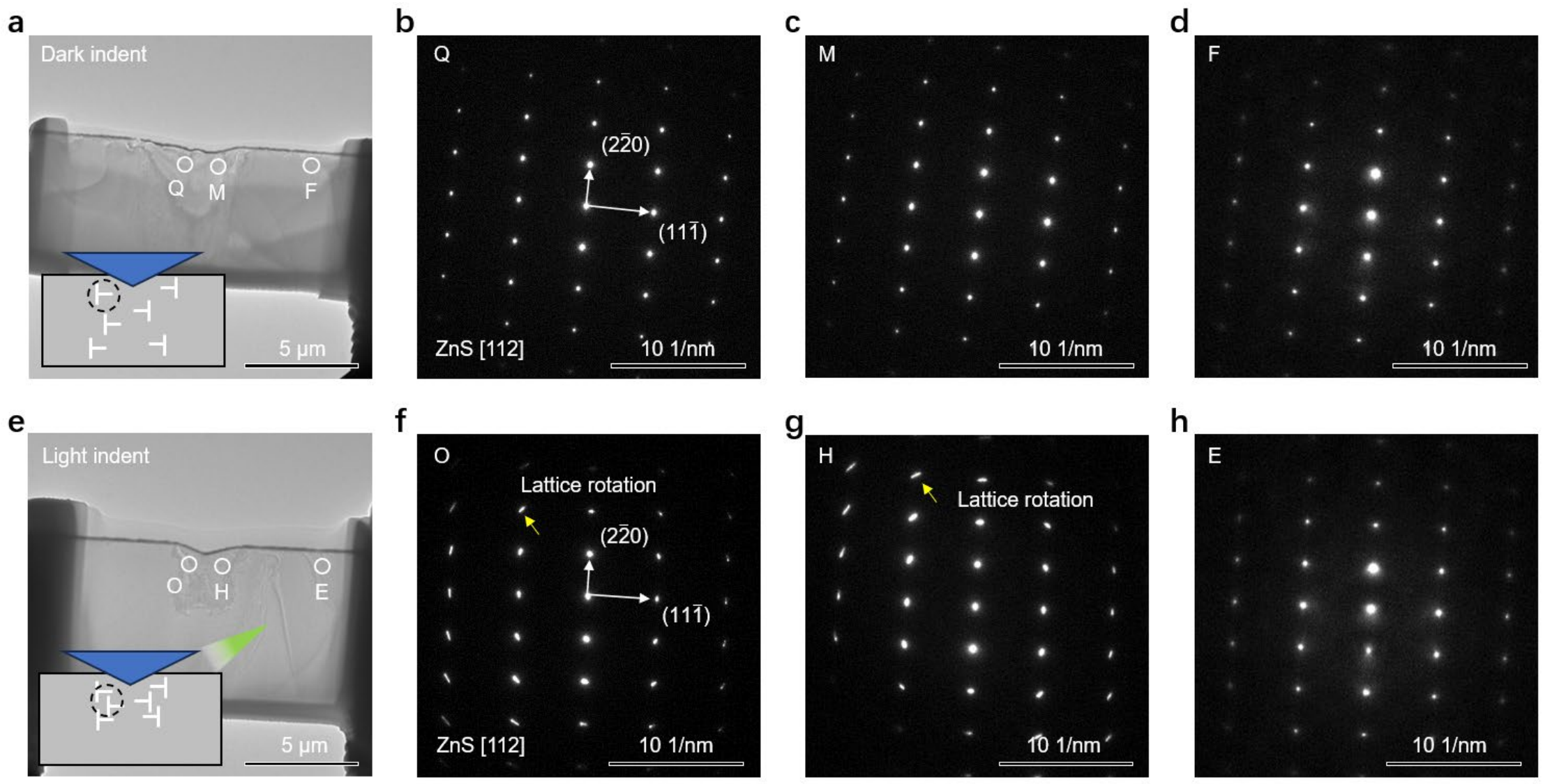

Fig. 3. Selected area electron diffraction (SAED) patterns of the samples underneath the dark and light indents. (a) An overview of the TEM sample underneath the dark indent and a schematic of the dislocation distribution. (b-d) SAED patterns corresponding to the marked areas in (a). An area away from the indent is selected as a reference. (e) An overview of the TEM sample underneath the light indent and a schematic of the dislocation distribution. (f-h) SAED patterns corresponding to the marked areas in (e). An area away from the indent is selected as a reference. The diffraction spots are split in (f) and (g), suggesting the lattice rotation.

Dislocations in ZnS generally dissociate into 30° and 90° partial dislocations due to its low stacking fault energy [42]. Because the 30° S partial dislocation possesses the lowest glide barrier [9, 23], here, we mainly focus on the mobility of the 30° S partial dislocation. Fig. 4 a and b present optimized models of 30° S dislocations at the ground state (in darkness) and the excited state (under light), respectively. Fig. 4b shows the photo-excited charges modified the atomic structure of the 30° S dislocation core and reveals that, particularly, the Zn-S bond angle changes under light. To keep consistent with experiments, we compare the dislocation evolution under the same strain in the darkness and illumination conditions with a relatively large model (~half a million atoms) in MD simulations (Fig. 4c and Fig. S9). The dipole atomic model contains a 30° S core and a 30° Zn core. The 30° Zn core possesses a higher glide barrier [9] and remains at its location during the glide of the 30° S core dislocation. As shown in Fig.

4d, the stress-time curves of both the ground state and excited states reveal that noticeable jumps represent the onset of dislocation motion when the stress is just above the Peierls stress (the strain is identical for the ground state and excited states). The dislocations in the excited state start to move at higher shear stress, which agrees with the increased hardness in our experiment (Fig. 1c). The Peierls stress is 1.38 GPa in the ground state and 1.64 GPa in the excited state (Fig. 4e). Fig. 4f shows that the dislocation in the excited state moves a shorter distance than that in the ground state, for example, 2579.82 Å in the excited state versus 2686.98 Å in the ground state at 400 ps, which qualitatively agrees with the experimental observation (Fig. 2). At the fixed strain, the difference in glide distance becomes larger with increasing time as shown in Fig. S10.

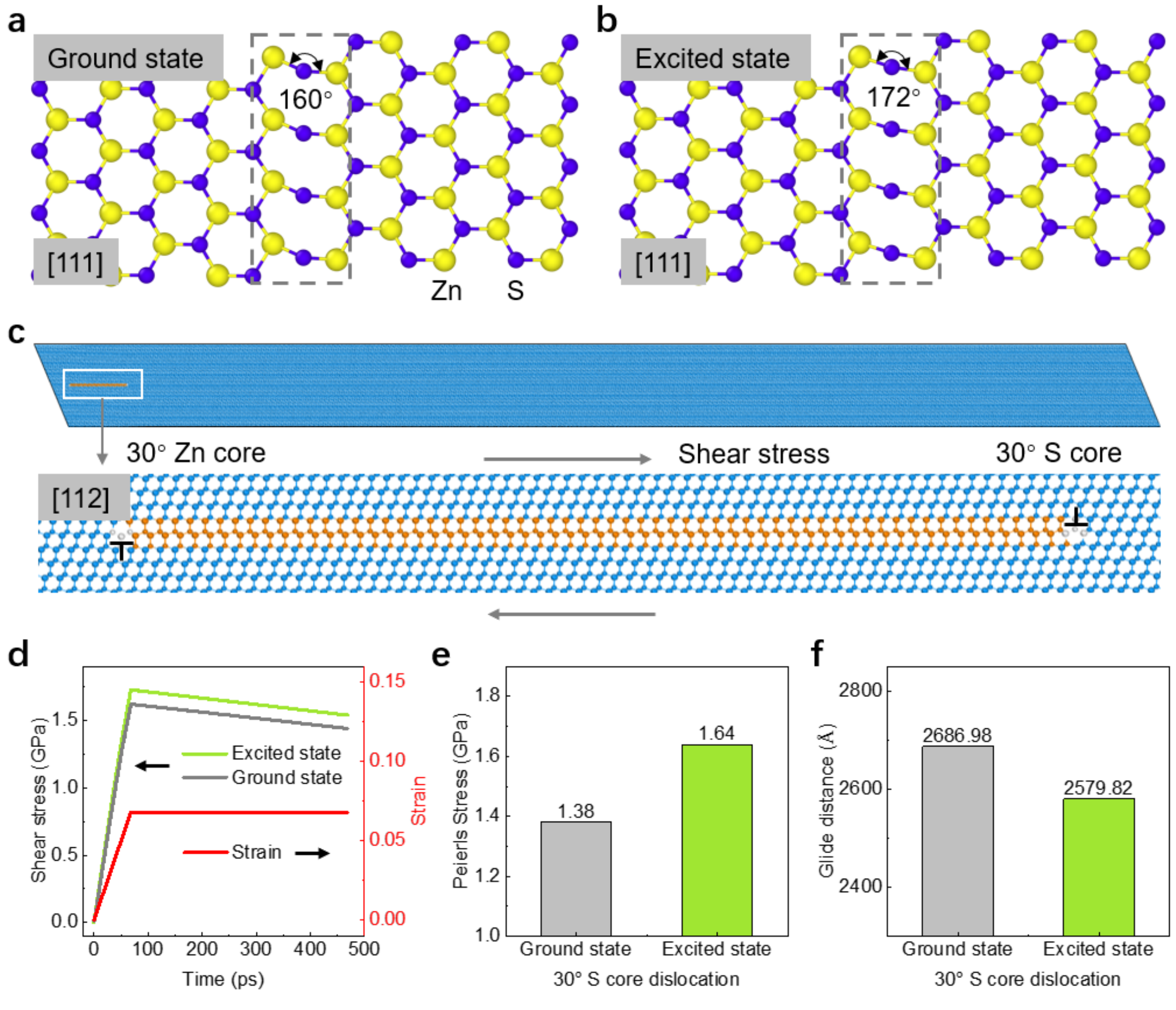

Fig. 4. MD simulations of reduced dislocation mobility under illumination. DFT-optimized atomic structure of a 30° S partial dislocation at the ground state (in darkness) (a) and excited state (under light) (b). (c) Atomic model to study the dislocation dynamics using MD simulations. The 30˚ S partial dislocation moves under shear stress, while the 30˚ Zn partial dislocation remains in its original location due to a higher glide barrier [9]. (d) Shear stress and shear strain as the functions of time for a 30˚ S partial dislocation in the ground and excited states. The strain is identical in both darkness and light, as the strain rate and depth in nanoindentation experiments. (e) The Peierls stress of the 30° S partial dislocation at the ground state and excited state. (f) The gliding

distances of the 30° S partial dislocations at the ground state and excited states at the same strain.

During plastic deformation, a high density of dislocations is generated, as illustrated in Fig. 2 and reported in the literature [35, 43]. The interaction between dislocations also significantly contributes to the mechanical response, such as the work-hardening effect [44, 45]. However, studies on the photo-plastic effect in semiconductors have primarily focused on isolated dislocations [23, 27]. Although these studies provided valuable insights into the influence of light on isolated dislocations, the interaction between dislocations is usually neglected. To address this, we investigated the effect of the excitation state on the elastic field of 30° partial dislocations, thereby elucidating how dislocation interaction changes under light. Fig. 5 shows the stress distribution around 30° S and 30° Zn partial dislocations in ZnS in the ground and excited states. The stress around the dislocation core under the excited state is higher than that under the ground state, as shown in Fig. 5c-d and Fig. S11. The stress distributions at the dislocation core through the range of our model (~ 80 Å) are provided in Figs. S12-13. The enhanced stress under the light condition suggests a stronger interaction between dislocations than those under the dark condition. In addition, the reduced dislocation mobility under light may lead to a higher dislocation density beneath the sample surface than that in darkness. Hence, compared to the dark condition, more dislocation junctions are likely to form under the light, impeding dislocation motion and further contributing to the hardening effect. Furthermore, the increased number of junctions may increase the roughness of dislocation lines, resulting in changes to the critical resolved shear stress of the dislocations [9]. Previous studies also reported that light could change the charge state [13, 25] and strain state [18] of point defects, which results in the changes in plastic and elastic properties of semiconductors.

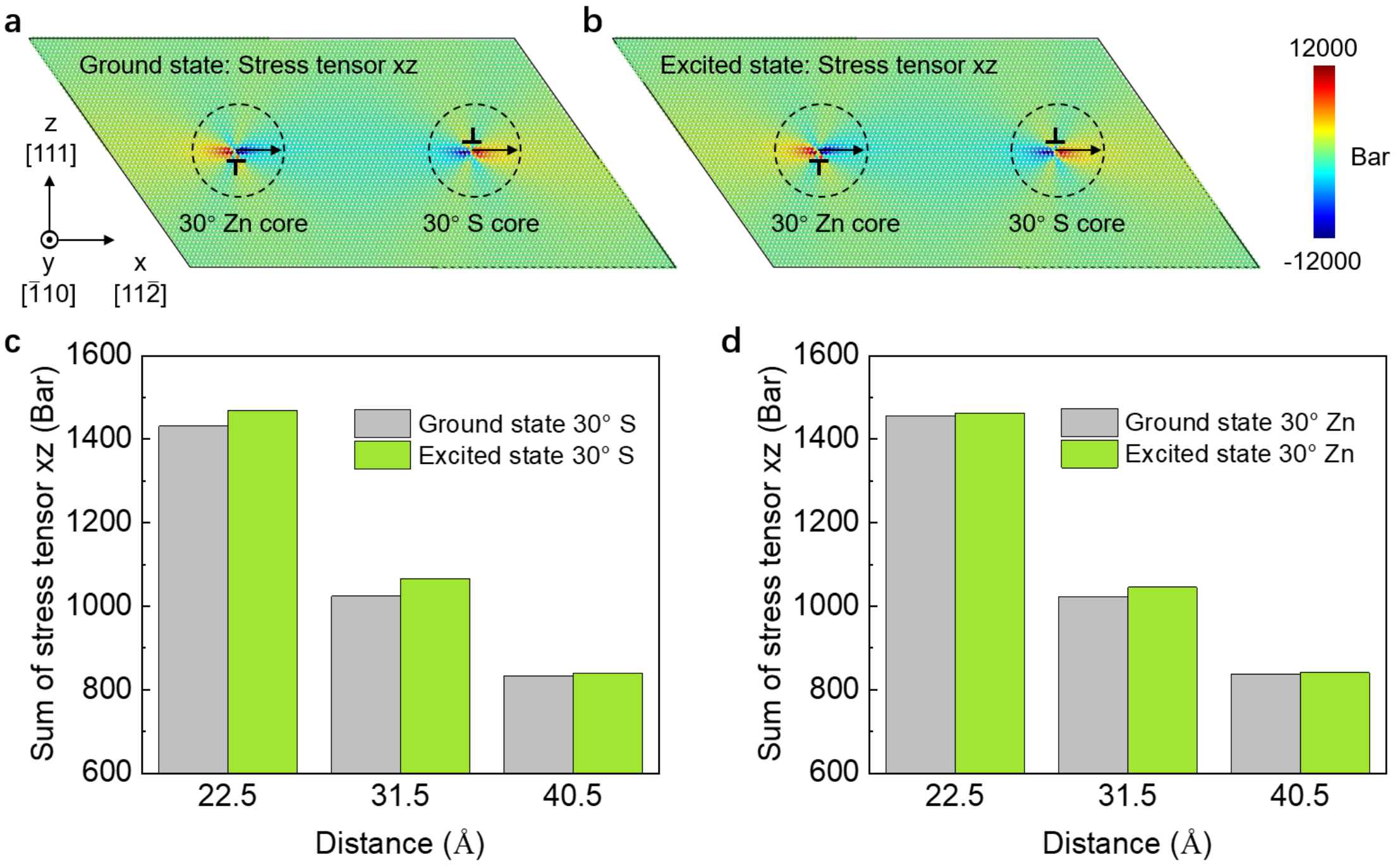

Fig. 5. Stress distribution around 30° S and 30° Zn partial dislocations in ZnS in the ground (dark) and excited (light) states: (a) ground state and (b) excited state. (c) The sum of the absolute value of the stress tensor xz as a function of distance away from the 30° S dislocation core. (d) The sum of the absolute value of the stress tensor xz as a function of distance away from the 30° Zn dislocation core.

The above results demonstrate that the light with a wavelength corresponding to the bandgap energy increases the hardness of the ZnS at the highest level, suggesting the reduction in dislocation mobility is associated with photo-induced electron-hole pairs. Previous studies reported the "after-effect" that is the stress is gradually back to the dark level when removing the light after several hundred seconds; Such effect may arise from the lifetime of photo-induced electrons and holes [36, 46, 47], further suggesting that the observed photo-plastic effect is attributed to the interaction between dislocations and photo-induced charge carriers. Hence, to simulate the photo-plasticity effect, we add electron-hole pairs into the semiconductors with electrically charged dislocations [48, 49]. For example, the S dislocation core is negatively charged in ZnS, while the Zn dislocation core is positively charged. Consequently, photo-excited holes are preferentially trapped in the 30° S partial dislocations. Our experimental evidence and atomic simulations offer valuable insights into the dislocation dynamics of semiconductors under light and in darkness, opening new opportunities for tuning their dislocation behavior and improving dislocation-related properties.

**Declaration of Complete Interests**

The authors declare that they have no known competing interests or personal relationships that could have appeared to influence the work reported in this paper.

**Acknowledgements**

M.L. and Y.Z. acknowledge the financial support from the Natural Sciences and Engineering Research Council of Canada Discovery (Grant no. RGPIN-2018–05731), NSERC Alliance International Collaboration grant (ALLRP 598774–24), Digital Research Alliance of Canada; the Canadian Foundation for Innovation, John R. Evans Leaders Fund (JELF) 38044; K.L. and Q.A. were supported by the National Science Foundation with funding number 2347218. M.L. thanks J. Xu at the Electron Microscopy Laboratory of Peking University for the support of the FIB system.

**Author contributions**

M.L. designed and carried out the nanoindentation, FIB and TEM experiments, analyzed the data and prepared the paper draft under the supervision of Y.Z. X.M provided guidance on the operation of TEM for dark-field images and electron diffraction patterns, and the analysis of TEM data. K.L. carried out MD calculations under the supervision of Q.A. B.K. performed AFM imaging of indents under the supervision of T.F. All authors contributed to this work through useful discussion, revision and comments to the paper.

**Supplementary Information**

# Direct evidence and atomic-scale mechanisms of reduced dislocation mobility in ZnS under illumination

Mingqiang Li[a,#], Kun Luo[b,#], Xiumei Ma[c], Boran Kumral[d], Peng Gao[c], Tobin Filleter[d], Qi An[b*], Yu Zou[a*]

[a] Department of Materials Science and Engineering, University of Toronto, Toronto, ON M5S 3E4, Canada

[b] Department of Materials Science and Engineering, Iowa State University, Ames, IA 50011, USA

[c] Electron Microscopy Laboratory, School of Physics, Peking University, Beijing, 100871, China.

[d] Department of Mechanical & Industrial Engineering, University of Toronto, Toronto, M5S 3G8, Canada

# These authors contribute equally to this work.

* Corresponding authors: mse.zou@utoronto.ca (Y.Z.); qan@iastate.edu (Q.A.)

## Materials and methods

### Materials

The single-crystalline ZnS (from MTI Corp.) for nanoindentation has a dimension of 10 mm × 10 mm × 1.0 mm with an orientation of <110>. The surface was optically polished. The resistivity is larger than $10^6$ ohm/cm.

### Nanoindentation

Nanoindentation tests were performed using a KLA iMicro system with a Berkovich diamond tip (Synton-MDP, Switzerland) at room temperature. During all tests, ambient light was blocked to study the effect of the applied illumination. The strain rate was set to 0.2 $s^{-1}$ with a hold time of 1.0 s. The target drift was maintained at 0.1 nm/s. The unloading time is set as 1.0 s for all the tests. A 90-second drift hold segment during unloading is turned on to measure the drift rate for calibration. Tests started only when the sample drift was below this threshold. We calibrated the nanoindentation system using a standard fused silica specimen before each experiment. Multiple tests were repeated for each sample to ensure reliable results. During the tests, the sample was mounted on an aluminum stage with a diameter of ~30 mm. The aluminum stage, with good thermal conductivity, maintained a stable sample temperature. The temperature was 25-30°C during the tests. The indentation depth is set to 600 nm. We developed an illumination system with two LEDs with a wavelength of 365 nm (corresponding to the bandgap of ZnS, ~ 3.5 eV).

### FIB and TEM

The TEM samples were prepared using the focused ion beam (FIB) technique (ThermalFisher Helios G4 UX). Fig. S5 illustrates the typical workflow for preparing cross-sectional TEM samples from indents. Each TEM sample is cut around the center of the indents. The surface morphology of TEM samples may vary depending on the thinning process. The ZnS was milled using a gallium ion beam at 30 kV with a beam current from 49 nA to 7 pA. Lower accelerating voltages of 5 kV and 2 kV with a beam current from 20 to 5 pA were used for further thinning to reduce the ion beam damage. Then, a 0.5-1 kV ion beam was used for the fine milling of the specimens. TEM images and electron diffraction patterns were obtained using the FEI Tecnai F20 microscope operated at 200 kV.

### AFM

Surface topology measurements of the indents were performed using an Asylum Cypher Atomic Force Microscope (Oxford Instruments, Santa Barbara, USA), operated in tapping mode with a silicon tip containing a Ti/Ir coating (Oxford Instruments ASYELEC.02-R2). Each indent was imaged across 12 × 12 μm scan regions with 512 pixels at a scan rate of 1 Hz.

### MD simulations

To simulate the influence of different light illumination conditions on dislocation motion, we performed MD simulations using machine learning potentials (MLPs) derived from density functional theory (DFT) and constrained DFT, which have been demonstrated to describe lattice parameters, elastic constants, and dislocation structures in ZnS under darkness and illumination [1]. Dislocation models were generated using Atomsk [2], resulting in a triclinic simulation cell containing a dislocation dipole positioned nearly equidistant on the (111) planes of ZnS. Within the cell, lattice vectors (a, b, c) are aligned along $\langle\bar{1}\bar{1}2\rangle$, $\langle1\bar{1}0\rangle$, and $\langle\bar{1}10\rangle$ to create a quadrupolar arrangement for dislocations that minimizes elastic interactions. The dislocation line is oriented along the b-axis on the (111) slip plane. For the 30° partial dislocations, a Burgers vector of $1/6\langle1\bar{2}1\rangle$ is used. The polar (111) planes in ZnS cause nonstoichiometry at the dislocation cores, yielding distinct S-core and Zn-core configurations. A dipole of straight glide-set 30° S-core and Zn-core dislocations with opposite Burgers vectors was placed along the a-axis (see Fig. 4c), naturally forming a stacking fault. The initial core structures were constructed as single-period (SP) configurations since core reconstructions are not favorable for 30° partial dislocations in ZnS [1]. The atomic model for simulating dislocation motion had dimensions of approximately 394.4 nm × 1.51 nm × 18.5 nm (comprising 576,000 atoms), ensuring a sufficiently long travel distance for the dislocations (Fig. 4c).

All MD simulations were conducted using LAMMPS [3, 4] with full periodic boundary conditions to avoid surface effects and excess elastic interactions between dislocations in replicated supercells. For determining dislocation travel distances, shear strains were applied along the Burgers vector on the (111) plane at 0.1 K. Recognizing that MD strain rates are orders of magnitude higher than those in experiments, we applied a shear strain rate of $1 \times 10^9$/s until a shear strain of 0.068 was reached—this being the minimum strain required to initiate dislocation motion in both the ground and excited states. Once this strain level was attained, the strain was held constant (as shown in Fig. 4d) to allow the system to relax. This quasi-static approach aims to approximate the equilibrium strain state observed under experimental plastic deformation, thereby mitigating dynamic effects associated with high strain rates. During the shear deformation, an NPT ensemble was applied with a barostat to maintain the zero external stress in all dimensions except the shear direction, ensuring ideal shear conditions. For visualization and analysis of the simulation outcomes, we utilized the Open Visualization Tool (OVITO) [5].

Additionally, we investigated the influence of various excitation states on the elastic field of 30° partial dislocations, thereby elucidating how dislocation–dislocation interactions affect dislocation mobility. The atomic model used for molecular mechanics (MM) simulations measured approximately 39.4 nm × 1.51 nm × 18.5 nm and consisted of 57,600 atoms, ensuring that the distance between dipole dislocation cores was around 20 nm [1]. Equilibrium MM simulations using LAMMPS provided the six stress tensor components ($\sigma_{xx}$, $\sigma_{\gamma\gamma}$, $\sigma_{zz}$, $\sigma_{xy}$, $\sigma_{xz}$, and $\sigma_{yz}$) for each atom. Atomic volumes were computed using the Voronoi tessellation method, and spatial averaging was performed to mitigate the differences between the local stresses of neighboring Zn and S atoms, thereby facilitating the analysis of stress distribution variations for the distance from the dislocation line. For the analysis of the spatial variation of stress as a function of the radial distance r (in the xz plane), a bin width of 3 Å was chosen. Moreover, based on the Peach–Koehler force theory, we placed particular emphasis on the $\sigma_{xz}$ and $\sigma_{yz}$ components, representing the glide component of the dislocation force.

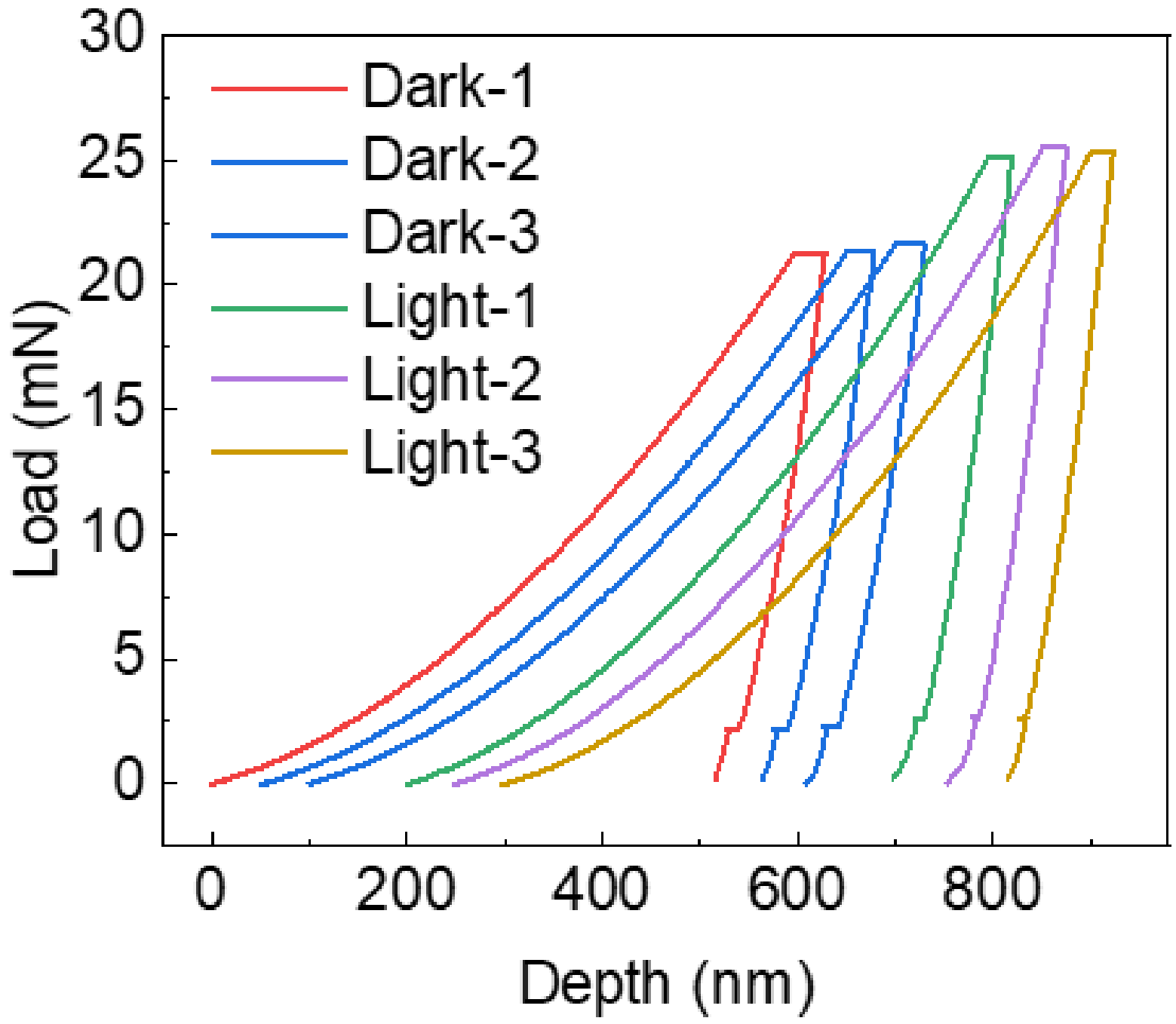

**Figure S1**. Repeating nanoindentation experiments showing the hardening effect of ZnS under 365 nm light. All the indents were performed on the {110} planes of ZnS. The curves are offset for clarity.

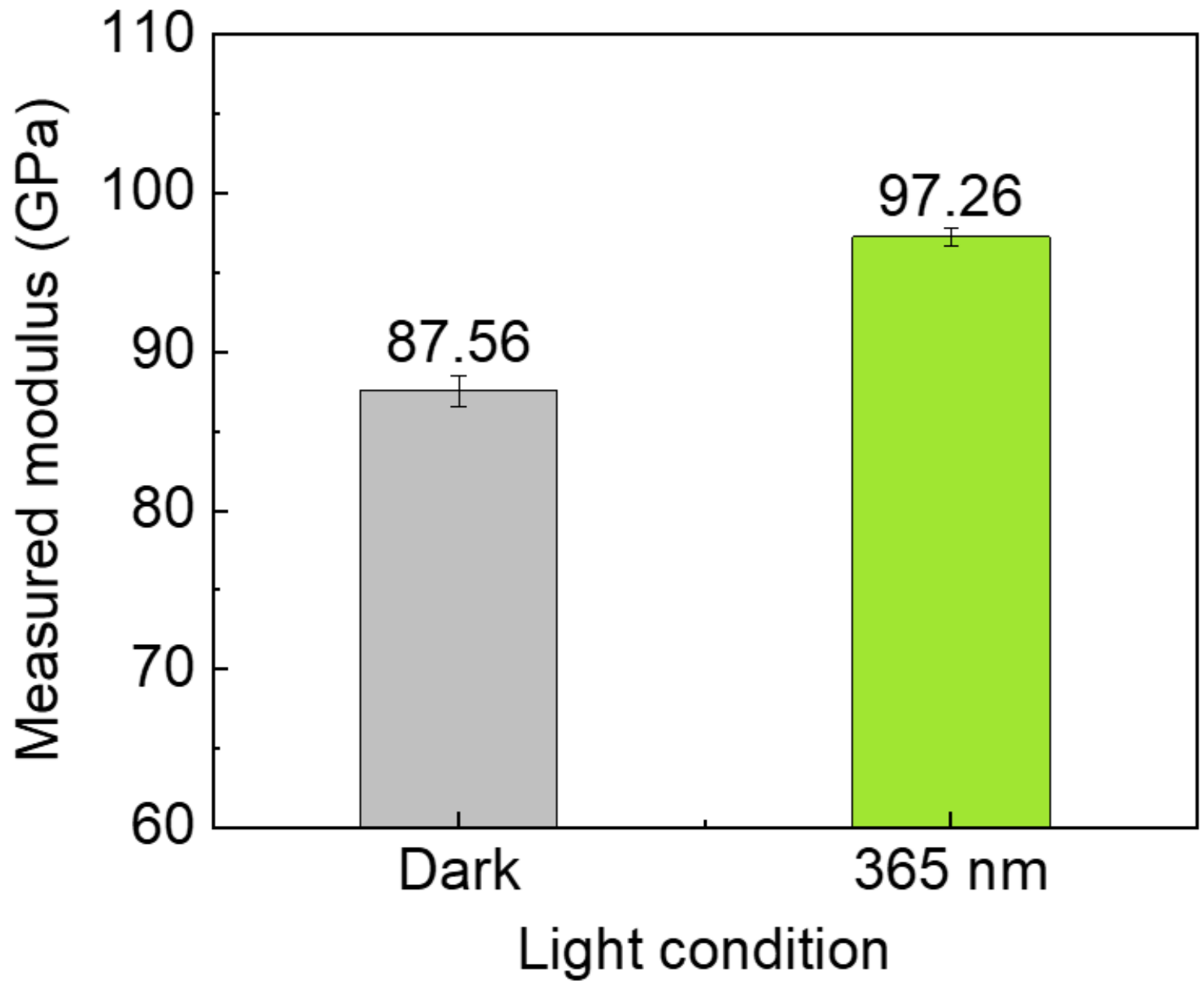

**Figure S2**. The measured modulus of ZnS in darkness and under illumination.

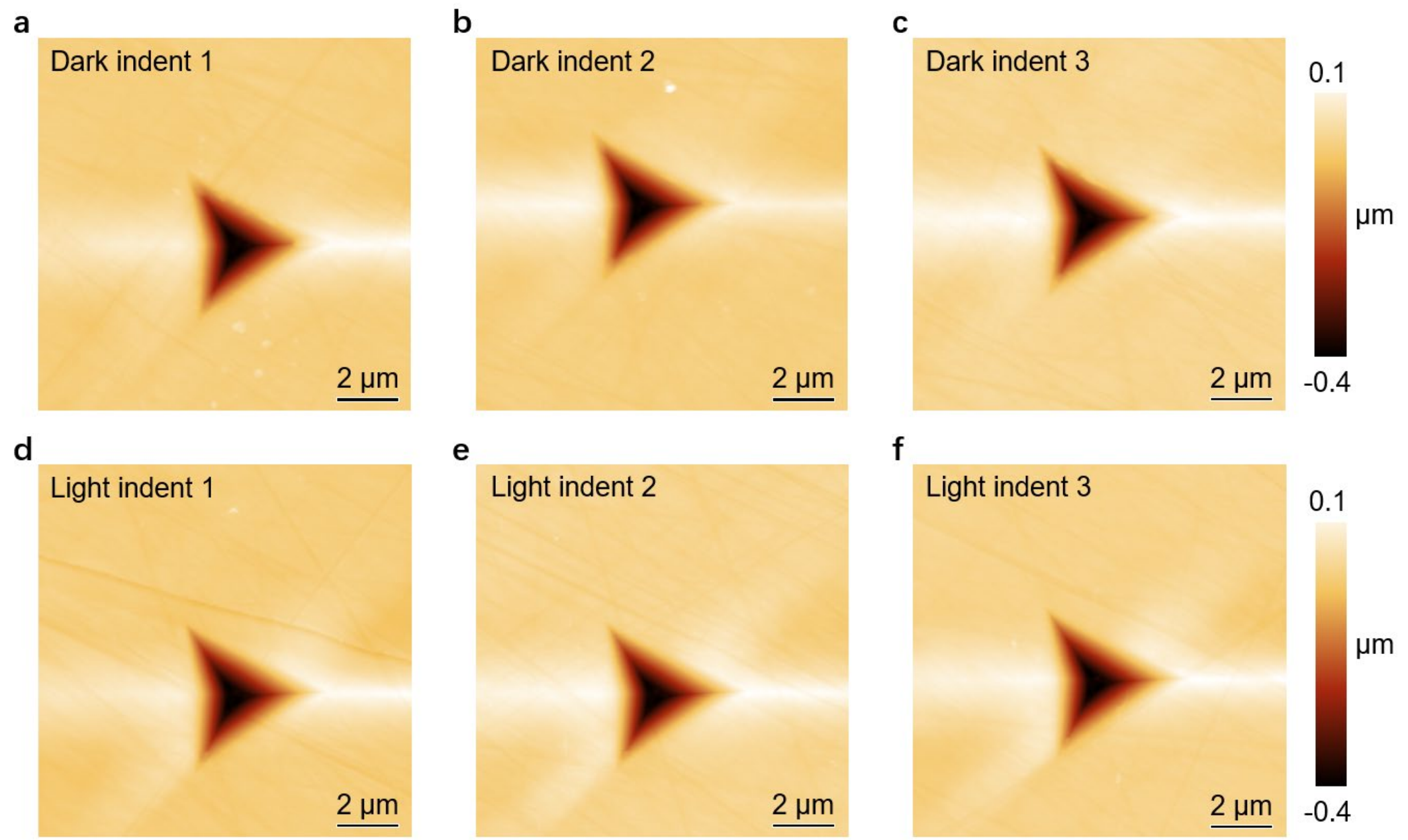

**Figure S3**. Topographic AFM images of indents on the ZnS surfaces. (a-c) Three topographic AFM images of indents in the darkness on ZnS. (d-f) Three topographic AFM images of the indents on ZnS under light. No micro-cracks are observed in both dark and light indents.

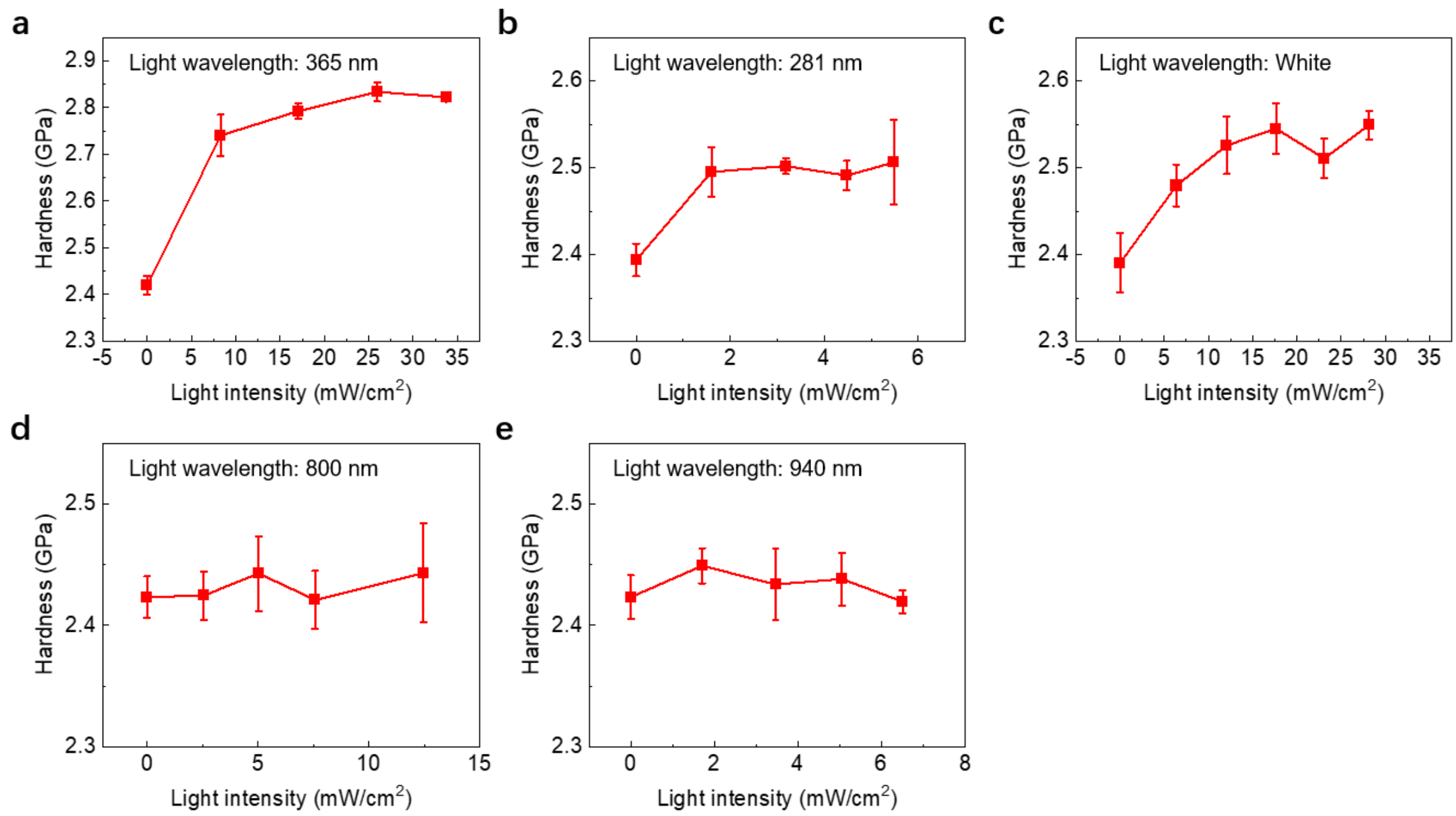

**Figure S4**. Dependence of the photo-plastic effect of ZnS on the light intensity and wavelength. (a-e) The hardness of ZnS under various light intensities and light wavelengths. The wavelength of light includes 365 nm (a), 281 nm (b), white (400-600 nm) (c), 800 nm (d) and 940 nm (e). The maximum effect occurs when illuminating the 365 nm light, corresponding to the intrinsic absorption edge of ZnS.

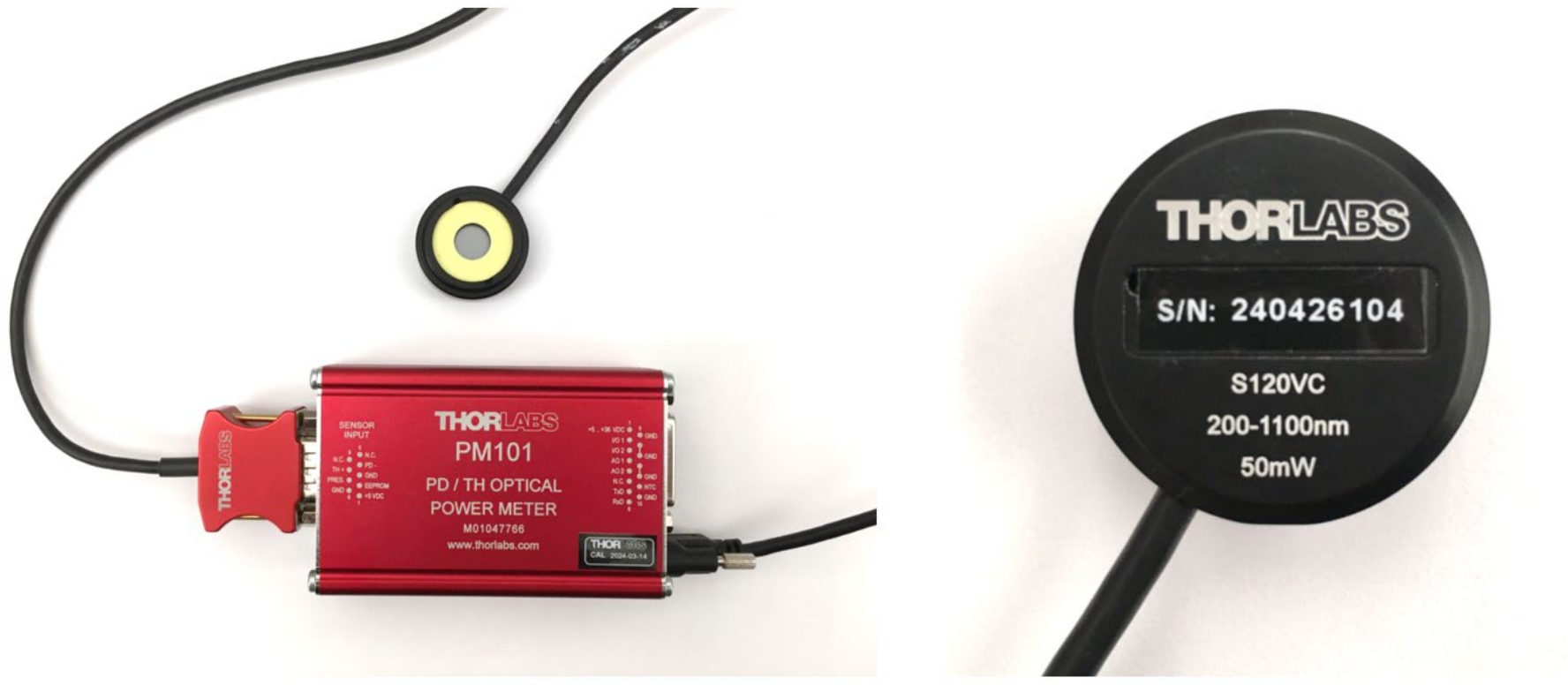

**Figure S5**. Optical power meter with a range of 200-1000 nm.

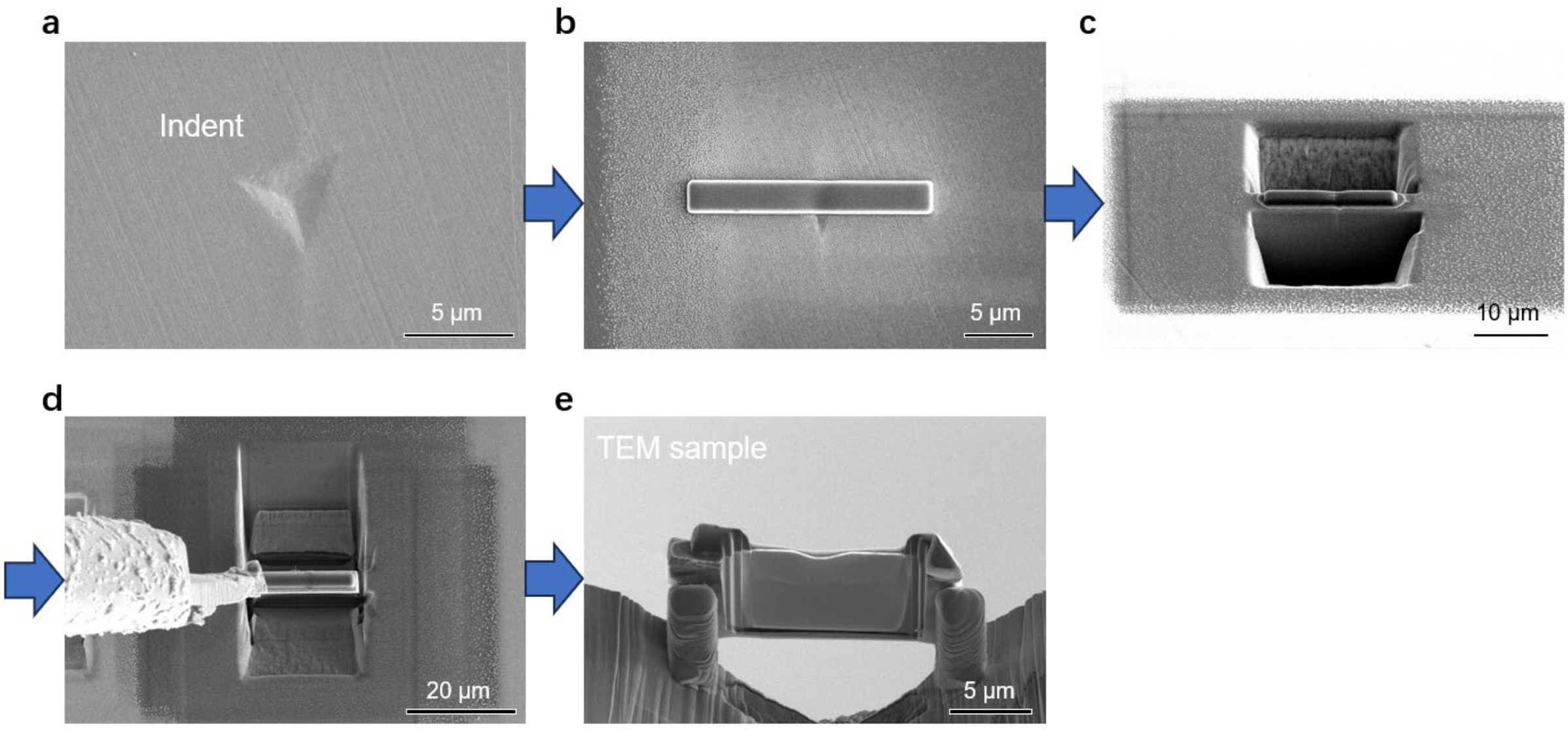

**Figure S6**. (a-e) Typical workflow of focused ion beam (FIB) fabrication of cross-sectional TEM samples.

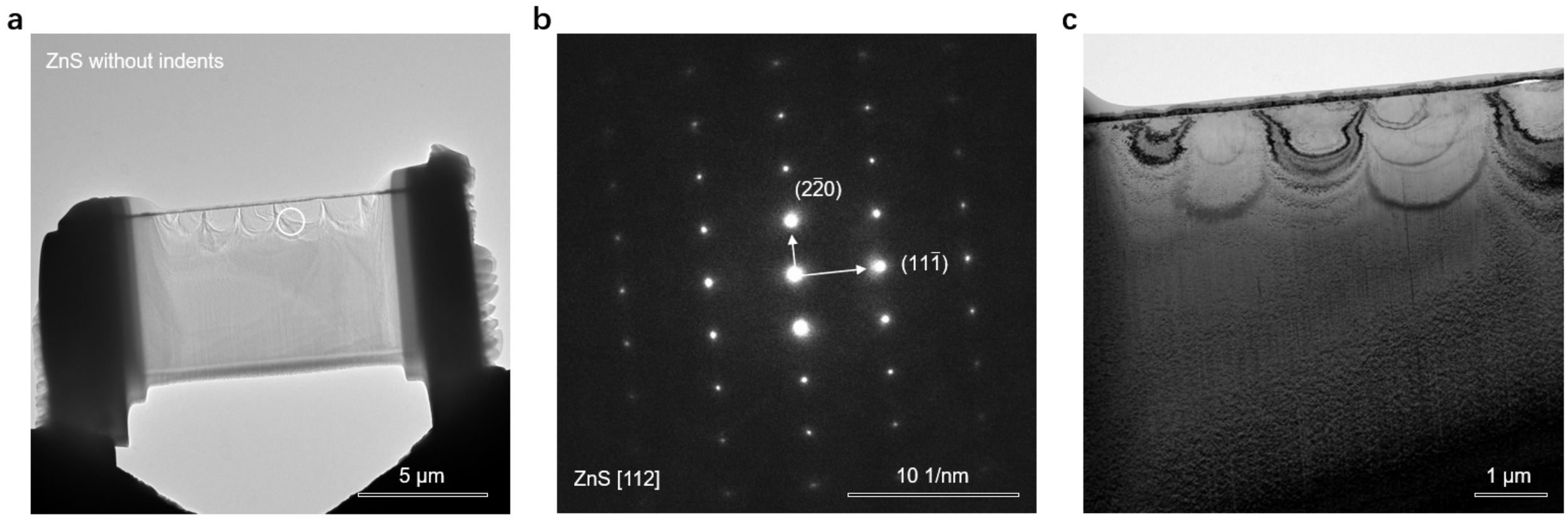

**Figure S7**. TEM characterization of the region of the sample without indents. (a) Overview of the cross-sectional TEM sample from the surface. (b) Electron diffraction pattern showing a single-crystalline feature. (c) Bright-field TEM images of the sample, suggesting no dislocation.

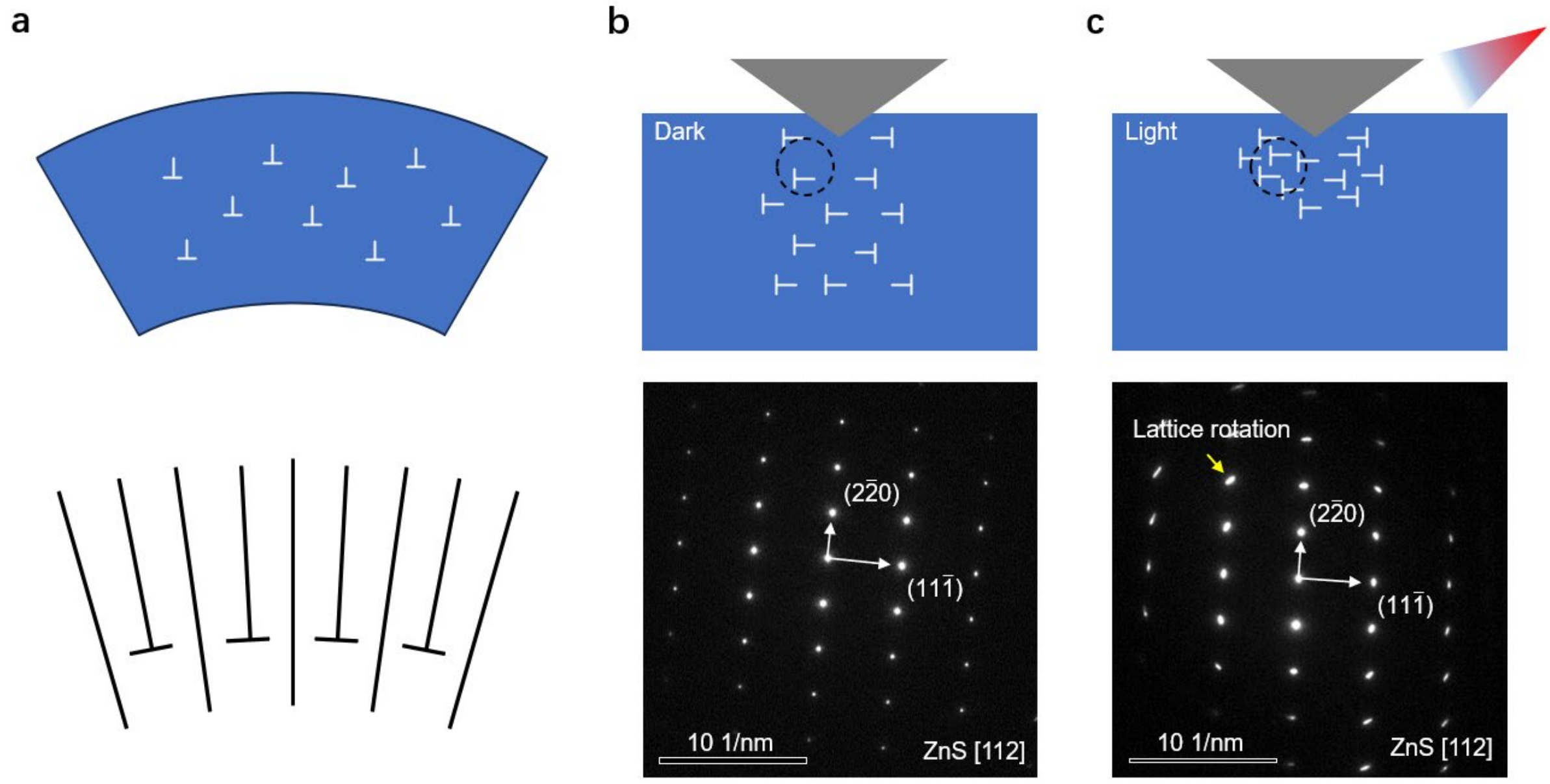

**Figure S8**. Lattice rotation under light indentation induced by dislocations. (a) Schematic diagram of the lattice curvature induced by distributions in the Nix-Gao model [6]. (b) Dislocation distribution in dark indent. The corresponding SAED shows no lattice rotation. (c) Dislocation distribution in light indent. The corresponding SAED shows clear lattice rotation. The distribution of dislocations in (b) and (c) is exaggerated for clarity.

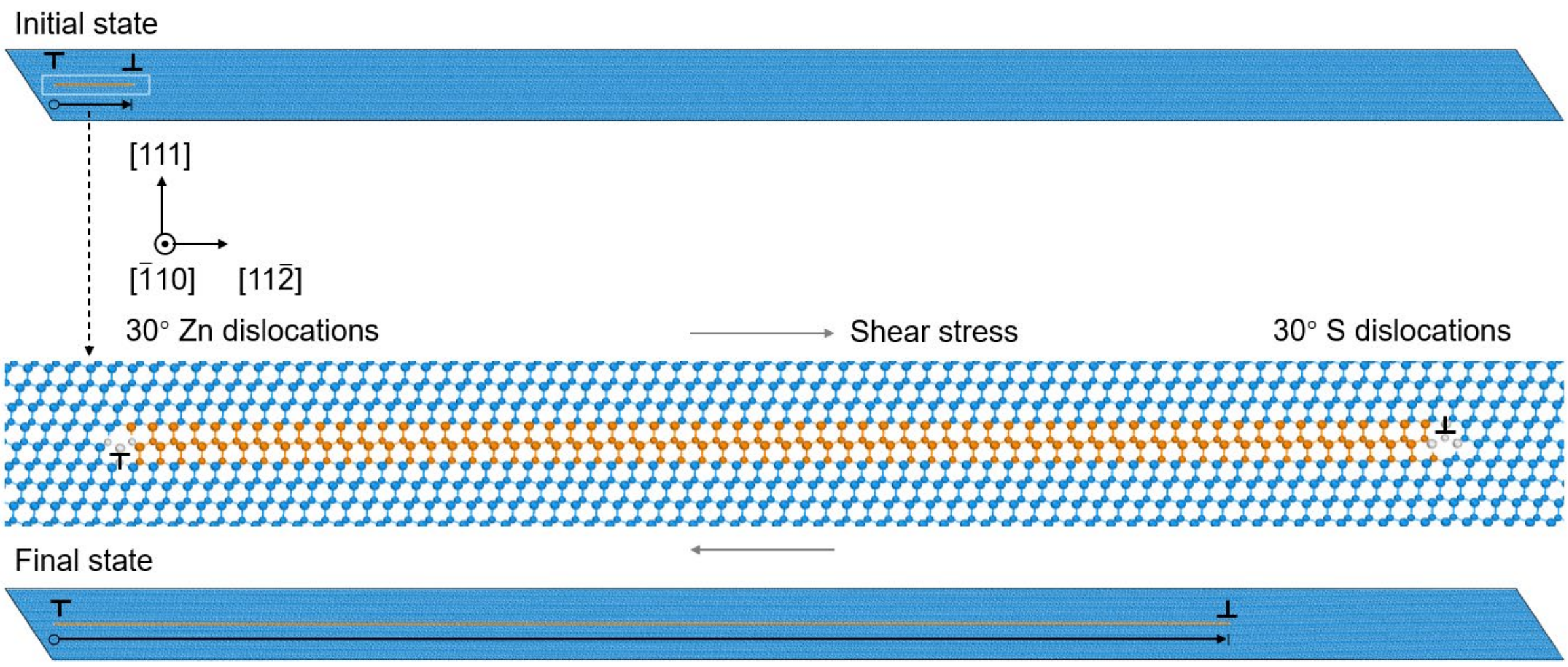

**Figure S9**. Molecular dynamics (MD) simulations of the dislocations glide. The 30˚ S partial dislocation moves under sheer stress while the 30˚ Zn partial dislocation remains its location due to a higher glide barrier.

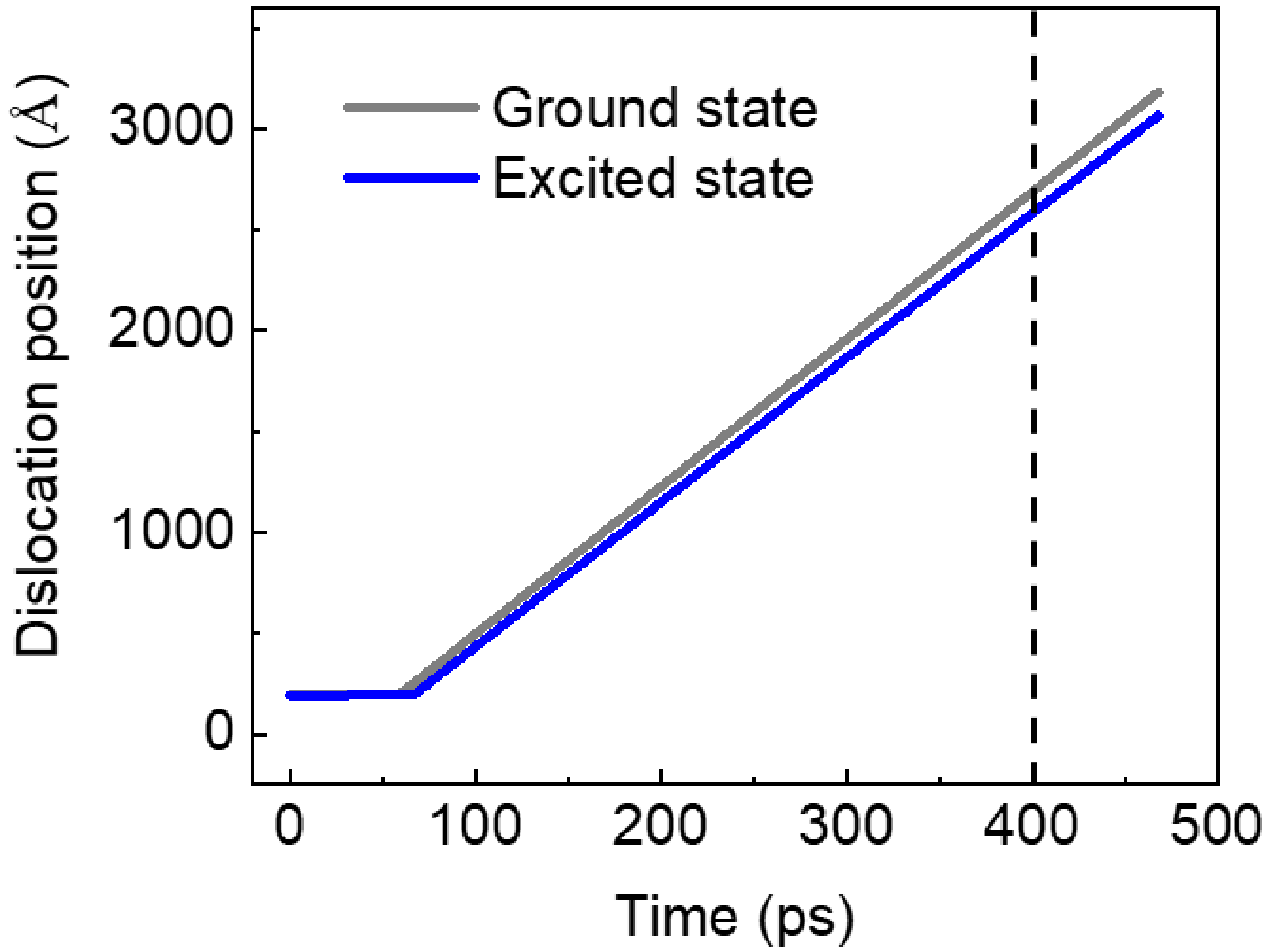

**Figure S10**. MD simulations tracking the position of the 30° S partial dislocation over time during shear deformation under both dark and light illumination conditions. The reference position (0) corresponds to the stationary 30° Zn partial dislocation. Notably, under illumination, the 30° S partial dislocation travels a shorter distance compared with its movement in darkness at equivalent strain levels (or the same time).

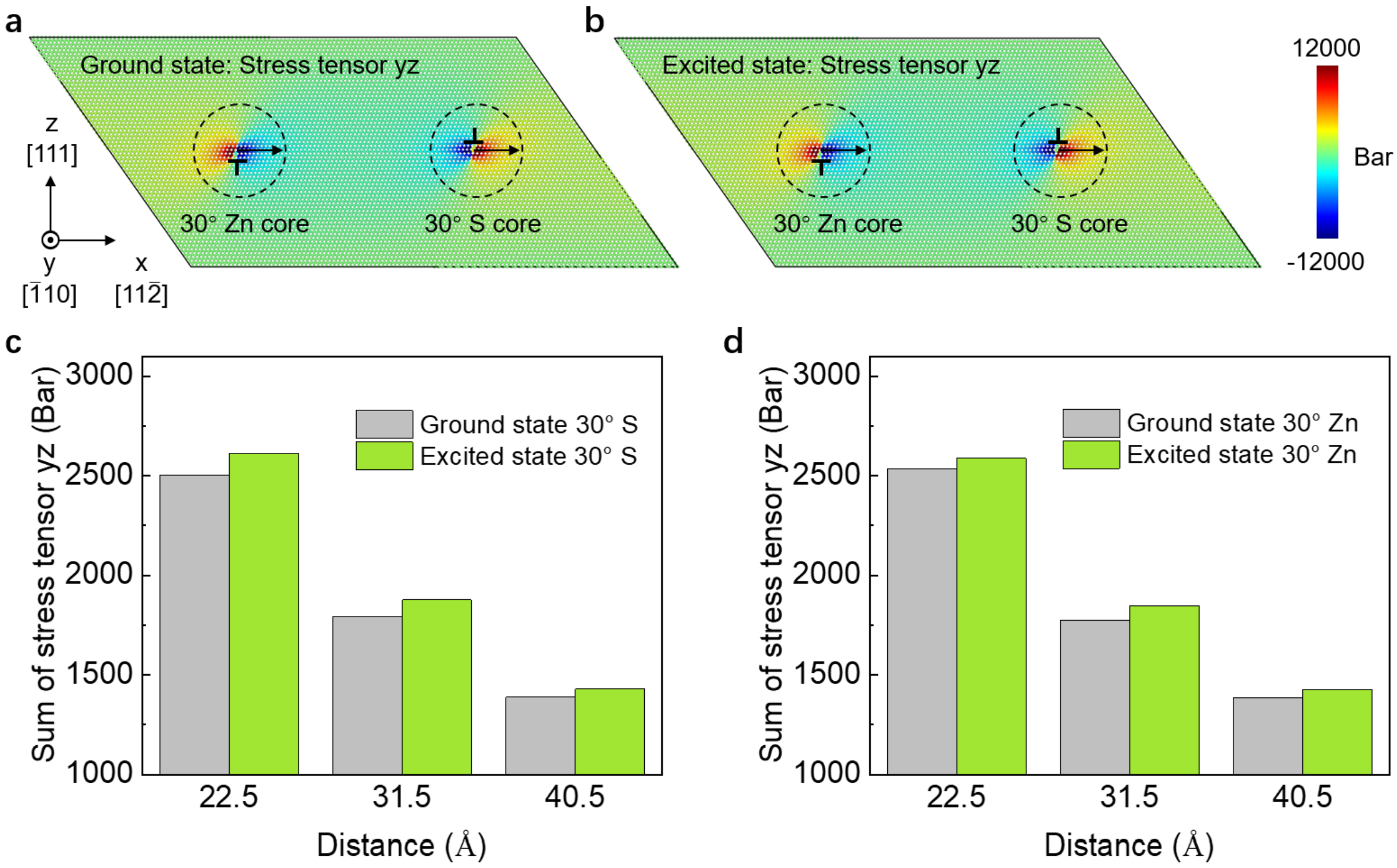

**Figure S11**. Stress distribution (stress tensor yz) around 30° S and 30° Zn partial dislocations in ZnS in the ground (dark) and excited (light) states: (a) ground state and (b) excited state. (c) The sum of the absolute value of the stress tensor yz as a function of distance away from the 30° S dislocation core. (d) The sum of the absolute value of the stress tensor yz as a function of distance away from the 30° Zn dislocation core.

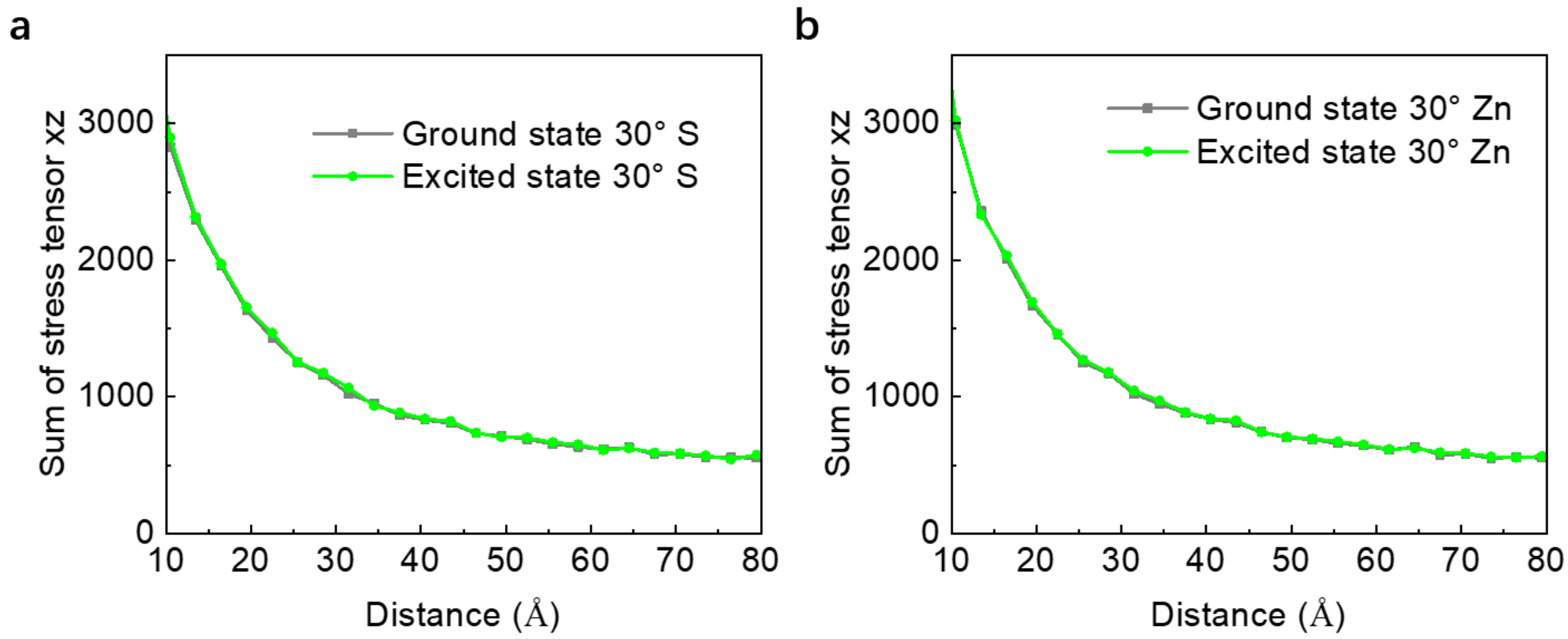

**Figure S12**. The stress tensor xz near dislocation cores. (a) The sum of the absolute value of the stress tensor xz as a function of distance away from a 30° S dislocation core. (b) The sum of the absolute value of the stress tensor xz as a function of distance away from a 30° Zn dislocation core. The difference between the ground state and the excited state is indistinct due to the wide range of stress values.

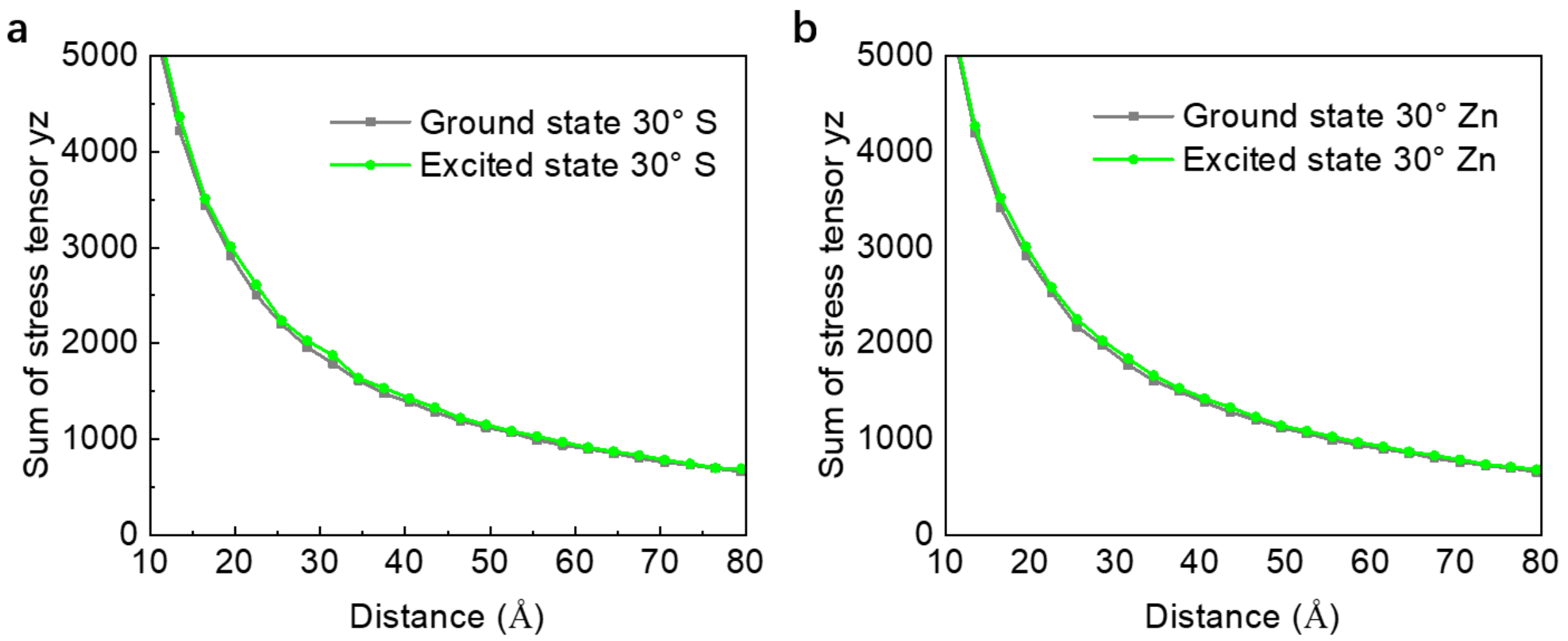

**Figure S13**. The stress tensor yz near dislocation cores. (a) The sum of the absolute value of the stress tensor yz as a function of distance away from a 30° S dislocation core. (b) The sum of the absolute value of the stress tensor yz as a function of distance away from a 30° Zn dislocation core. The difference between the ground state and the excited state is indistinct due to the wide range of stress values.